# Epitaxial Films as Model Platform for Understanding Compositionally Complex Electrocatalysts

*Satyakam Kar[a], Alejandro E. Perez-Mendoza[b], Huixin Xiu[c,d], Miran Joo[c], Kirill V. Yusenko[e], Ulrich Hagemann[f], Christoph Somsen[g], Janine Pfetzing-Micklich[h], Christina Scheu[c], Corina Andronescu[b,*], and Alfred Ludwig[a,h,i,*]*

[a] Chair for Materials Discovery and Interfaces, Institute for Materials, Ruhr University Bochum, Universitätsstraße 150, 44801 Bochum, Germany

[b] Chemical Technology III, Faculty of Chemistry and CENIDE, Center for Nanointegration, University of Duisburg-Essen, Universitätsstraße 7, 45141 Essen, Germany

[c] Max Planck Institute for Sustainable Materials, Max-Planck-Straße 1, 40237 Düsseldorf, Germany

[d] School of Materials and Chemistry, University of Shanghai for Science and Technology, 516 Jungong Road, Shanghai, 200093, China

[e] Institute of Geology, Mineralogy and Geophysics, Faculty of Geosciences, Ruhr-University Bochum, Universitätsstraße 150, 44801 Bochum, Germany

[f] Interdisciplinary Center for Analytics on the Nanoscale, University of Duisburg-Essen, Carl-Benz- Straße. 199, 47057 Duisburg, Germany

[g] Chair for Materials Science and Engineering, Institute for Materials, Ruhr University Bochum, Universitätsstraße 150, 44801 Bochum, Germany

[h] Center for Interface-Dominated High-Performance Materials (ZGH), Ruhr University Bochum, Universitätsstraße 150, 44801 Bochum, Germany

[i] Reseach Center Future Energy Materials and Systems (RC FEMS), Ruhr University Bochum, Universitätsstraße 150, 150, 44801 Bochum, Germany

*Corresponding authors, email: alfred.ludwig@rub.de , corina.andronescu@uni-due.de

**Abstract**

Compositionally complex solid solutions provide a unique route for engineering high-performance electrocatalysts, where the polyelemental surface composition can be seamlessly tuned to optimize activity, selectivity, and stability. However, the mechanistic understanding of these electrocatalysts remains limited by the lack of a model system with a crystallographically-defined surface that is compatible with correlative, multi-scale characterization. Here, we present epitaxial films as a model platform for studying compositionally complex electrocatalysts. Using magnetron sputtering, we realize (111) epitaxial Ir-Pd-Pt-Rh-Ru films on (0001) sapphire substrate via a (111) Pt buffer layer, confirmed by X-ray diffraction and transmission electron microscopy. The growth approach is applicable across a broad composition range and produces smooth surfaces (root mean square roughness < 1 nm) with micrometer-sized grains in the nanoscale films. With these films, we demonstrate direct structure-activity mapping at the nanoscale through precise co-localization using micro-indents and performing correlative atomic force microscopy, electron backscatter diffraction, and scanning electrochemical cell microscopy. Our work establishes a model platform for fundamental scale-bridging characterization and paves the way for rational design of compositionally complex electrocatalysts.

Keywords: Epitaxial Films, Compositionally Complex Materials, Electrocatalysis, Co-localized Characterization, Hydrogen Evolution Reaction

## 1. Introduction

Compositionally complex solid solution (CCSS) systems containing five or more principal elements (often referred to as high entropy alloys) offer new opportunities in material discovery and design.[1–4] Their vast multinary composition space enables exploration and tuning of material properties that are inaccessible in conventional material systems based on one or two principal elements. Recent advances in CCSS exploration have already produced materials with superior structural and functional properties.[5–10] In particular, CCSSs introduce a new paradigm for electrocatalyst design, enabling the design of novel catalyst materials that could be simultaneously active, selective, and stable owing to their polyelemental surface atom arrangements (SAAs).[11–13]

The surface of CCSS hosts abundant and diverse SAAs (Figure 1a), that can be tuned to achieve the desired multifunctionality. For instance, the activity of an electrochemical reaction can be improved by optimizing the adsorption energy of the reactant species on the catalyst surface. Since this energy is influenced by the local atomic environment, a continuous distribution of adsorption energies can be generated in CCSSs.[11–13] In a quinary CCSS catalyst with only on-top binding, the binding atom primarily determines the adsorption energy, which is modulated by neighboring atoms in the SAA. Accordingly, theoretical calculations indicated that five peaks appear in a near-continuous adsorption energy distribution (Figure 1a) corresponding to the five elements in the quinary CCSS.[12] By selecting an appropriate CCSS system and adjusting its composition, one has the means to tailor the adsorption energy distribution for optimum catalytic activity. Significant efforts have therefore been directed towards the development of CCSS electrocatalysts, which include advances in synthesis techniques,[14,15] high-throughput synthesis and characterization,[16–19] and machine learning accelerated theoretical predictions[12,20,21] to explain composition–structure–activity trends. However, these studies largely rely on nanoparticles and polycrystalline thin films, both of which contain heterogeneous crystallographic surfaces and hence are not ideal for fundamental investigations into SAAs. Nanoparticles, for instance, often have an irregular shape with multiple facets, making it difficult to isolate facet-dependent activity and catalytically relevant polyelemental SAAs.[11] While this aspect has caught recent attention,[22–24] it is still pertinent to develop a model system platform with well-defined surface that is accessible to a wide array of characterization techniques, and in particular, correlative multimodal characterization of the surface.[25–28]

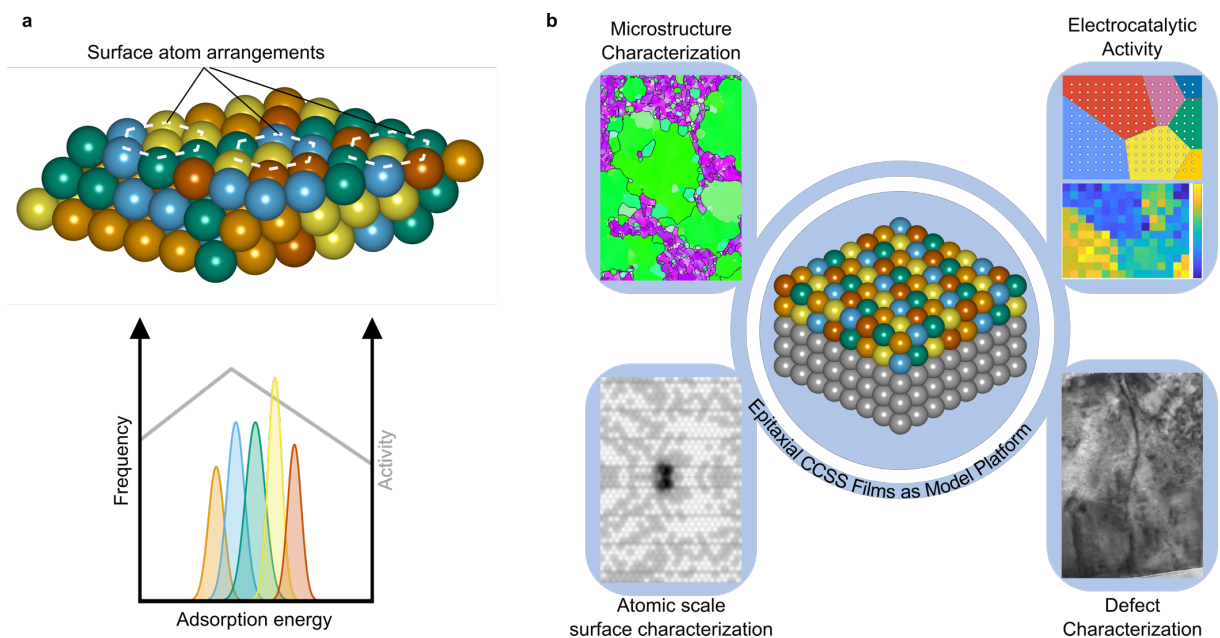

**Figure 1.** *Unique features of epitaxial CCSS films for electrocatalysis.* (a) Illustration depicting a model quinary (111) FCC film surface with diverse polyelemental surface atom arrangements, which result in a near continuous adsorption energy distribution of available binding sites for an electrocatalytic reaction.[12] Five characteristic peaks, corresponding to the five elements in the CCSS, are depicted here for the simplest case of on-top binding. This distribution can be tuned with composition changes to unlock optimum electrocatalyst performance. (b) Sketch illustrating the utility of epitaxial CCSS films as a model system for scale-bridging and correlative investigations conducted by EBSD, SECCM, TEM, and STM.

Epitaxial CCSS thin films offer such a model platform. Smooth, large-grained epitaxial films enable (a) microstructural analysis by electron backscatter diffraction (EBSD), (b) correlative investigation with locally measured electrocatalytic activity by scanning electrochemical cell microscopy (SECCM), (c) nanoscale defect characterization by (scanning) transmission electron microscopy ((S)TEM), and (d) atomic-resolution probing of surface using (electrochemical) scanning tunneling microscopy.[29] These capabilities make epitaxial CCSS films uniquely suited for scale-bridging investigations of promising CCSS compositions identified with high-throughput workflows. To date, epitaxial CCSS films have not been explored for electrocatalytic studies. Reports on epitaxial growth of CCSS films are also limited, for e.g., epitaxial growth of near equiatomic Co-Cr-Fe-Ni films by magnetron sputtering for studying microstructure and electronic properties,[30–32] and near equiatomic Dy-Gd-Ho-Lu-Tb films by molecular beam epitaxy for studying magnetic properties.[33] Hence, a general approach for epitaxial growth of CCSS across a broad compositional range and using elements relevant for electrocatalysis[12,14,24,34] remains to be established.

In this work, we use magnetron sputtering to demonstrate epitaxial (111) growth of Ir-Pd-Pt-Rh-Ru-based nanoscale CCSS films on *c*-plane sapphire ($Al_2O_3(0001)$) using a Pt buffer layer. By X-ray diffraction (XRD) and EBSD, we identify two epitaxial growth relationships that result in distinct microstructural characteristics on the surface. (S)TEM of the film cross-section reveals a coherent buffer-film interface and exotic defect structures that can occur in CCSS films. Furthermore, we implement a scale-bridging surface navigation protocol based on micro-indents to enable correlative nanoscale analysis of structure and electrocatalytic activity of the CCSS film by atomic force microscopy (AFM), EBSD, and SECCM. Together, these findings demonstrate that epitaxial films combined with on-film navigation structures serve as an effective platform for advancing mechanistic understanding and development of CCSS electrocatalysts.

## 2. Results and Discussion

The epitaxial growth of near equiatomic Ir-Pd-Pt-Rh-Ru-based-CCSS (hereafter referred to as CCSS) films begins with selecting a single crystal substrate that: (a) offers minimum lattice misfit with platinum group elements and (b) is chemically inert during hot deposition, typically necessary for epitaxial growth. The *c*-plane (0001) sapphire substrate, known for its success in epitaxy of noble metals,[35–38] was therefore chosen to realize epitaxial CCSS films by co-sputtering (see Methods, Supporting Information Table S1 for the deposition conditions and Table S2 for the EDS composition of films). Several approaches were pursued to achieve epitaxial growth: (a) direct deposition on the substrate, (b) two-step hot deposition, and (c) deposition with a Pt buffer layer. Figure 2a and b depict the surface topography of nominally 60 nm thick CCSS films deposited directly on $Al_2O_3$(0001) at 297 K (Figure 2a), and 873 K (Figure 2b). Both films exhibit granular surfaces, with grain size and roughness increasing with the increase in deposition temperature. Particularly, the root mean square (RMS) roughness increases from 0.83 nm at 297 K to 2.71 nm at 873 K. Using a two-step hot deposition process (15 nm film growth at 873 K followed by 45 nm film growth at 673 K) results also in a granular surface (Figure 2c), but with a lower RMS roughness of 1.53 nm than that obtained by single step hot deposition at 873 K. Besides these variations in surface roughness, X-ray diffractograms of the three films (Figure 2d) reveal polycrystalline growth with FCC crystal structure. This suggests that conventional hot deposition or multi-step deposition processes,[35,36,39,40] that led to smooth epitaxial growth of platinum group metals, do not work the same way for CCSS films. The underlying reason could be the chemical complexity of the incoming sputter flux, which has to grow epitaxially on the oxide substrate that offers lower adatom mobility due to poor thermal conductivity and worse surface wettability than metallic substrates.

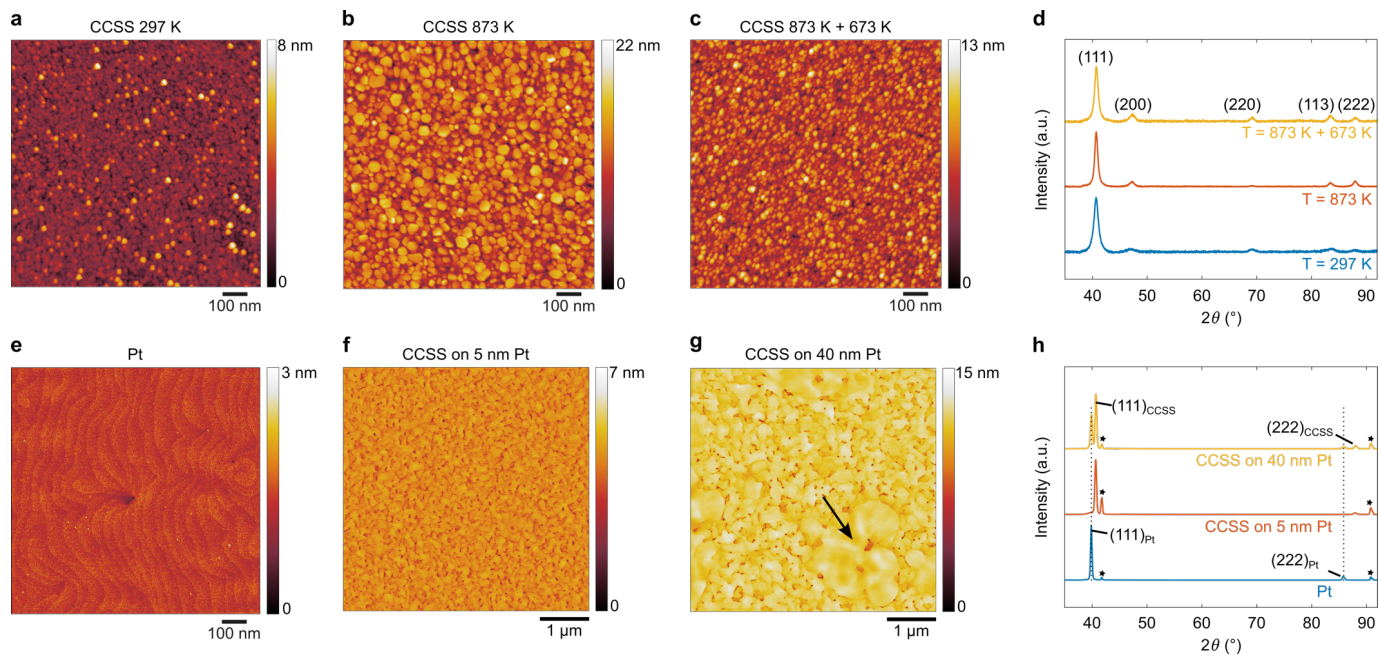

**Figure 2.** *Process design for smooth epitaxial CCSS films*. Surface topography of nominally 60 nm thick films grown on Al$_2$O$_3$(0001) at (a) 297 K, (b) 873 K, and (c) with two-step deposition at 873 K for 15 nm and then at 673 K for 45 nm. The films have a granular surface, with roughness increasing as the deposition temperature increases. (d) XRD of the three films reveal FCC crystal structure with polycrystalline growth. (e) Surface topography of a nominally 60 nm thick Pt film grown on Al$_2$O$_3$(0001) at 873 K. The film surface is smooth and continuous, appropriate for use as a buffer layer. (f) Surface topography of a nominally 60 nm thick CCSS film with 5 nm thick Pt buffer grown at 873 K. The CCSS film surface has flat grains with discontinuities, but smooth surface. (g) Surface topography of a nominally 100 nm thick CCSS film with 40 nm thick Pt buffer grown at 873 K. Roughness and grain size increase with increase in film thickness. Micrometer-sized grains are visible, as indicated by an arrow. (h) XRD of the Pt film and the CCSS films on Pt buffer show only (111) and (222) reflections with substrate reflections (★), indicating epitaxial growth.

To address these challenges, we explored the use of an elemental buffer layer, which grows epitaxially on Al$_2$O$_3$(0001). Pt was selected due to its robust epitaxy on Al$_2$O$_3$(0001), demonstrated across multiple deposition techniques.[35,36,40] The presence of Pt in the studied CCSS system also might help to minimize the lattice mismatch between the buffer layer and the CCSS film. Figure 2f shows the surface topography of a nominally 60 nm thick Pt film deposited on Al$_2$O$_3$(0001) at 873 K. The film surface is continuous and smooth (RMS roughness: 0.17 nm), with terraces observed as wiggly lines, confirming that the Pt buffer surface is ideal for growing smooth CCSS films. Hence, we deposited a nominally 60 nm thick CCSS film with a 5 nm thick Pt buffer at 873 K. The resulting CCSS film surface, depicted in Figure 2f, features discontinuities with flat grains fused together (see Supporting Information Figure S1a for a magnified view of the film surface). The measured RMS roughness of 0.36 nm is significantly (7.5 times) lower than the roughness of the CCSS film grown directly on Al$_2$O$_3$(0001) (Figure 2b), demonstrating the beneficial effect of the Pt buffer. To further understand the evolution of surface morphology and microstructure with increasing film thickness, we examined a nominally 100 nm thick CCSS film grown on 40 nm thick Pt buffer (Figure 2g, Supporting

Information Figure S1b). While the RMS roughness increased to 0.85 nm, the size of the desired flat grains has increased substantially, with some grains extending beyond 1 μm, as highlighted by the arrow. The X-ray diffractograms of these three aforementioned films (Figure 2h) exhibit only (111) and (222) reflections of the CCSS film as well as (111) and (222) reflections of Pt in the case of the pure Pt film sample and the CCSS film sample with 40 nm Pt buffer, indicating that the films grow epitaxially on $Al_2O_3$(0001).

To confirm epitaxial growth and determine the epitaxial relations, we measured {220} pole figures of the pure Pt film and the CCSS films grown with a Pt buffer. The {220} pole figure of the Pt film (Figure 3a) shows a set of three high intensity spots at sample tilt ($\psi$) of 35° and separated by 120° sample rotation ($\varphi$), consistent with the three-fold symmetry of the (111) oriented film surface. In addition, three low intensity spots mirroring the high intensity spot positions are observed at ($\psi = 35°$, $\varphi = 60°$, 180°, 300°), indicating {111} twinning of Pt during film growth. With respect to the substrate, the epitaxial growth relation is determined as Pt (111)[$\bar{1}$10] ∥ $Al_2O_3$(0001)[2$\bar{1}\bar{1}$0] (designated as OR2,[30] see Supporting Information Figure S2a for an illustrative sketch). The {220} pole figure of the CCSS film grown on 5 nm thick Pt buffer (Figure 3b) exhibits six high intensity spots at $\psi = 35°$, separated by 60° sample rotation. It implies that the CCSS film, like Pt, also undergoes {111} twinning during growth. All spots in the pole figure are rotated by 30° relative to Figure 3a, which suggests a different epitaxial growth relationship, determined as CCSS (111)[$\bar{1}$10] ∥ $Al_2O_3$(0001)[01$\bar{1}$0] (designated as OR1,[30] see Supporting Information Figure S2b for an illustrative sketch). For the CCSS film grown on 40 nm thick Pt buffer, the {220} pole figure (Figure 3c) has twelve spots at $\psi = 35°$, separated by a 30° sample rotation. Hence, this CCSS film exhibits both OR1 and OR2 growth relations alongside {111} twinning, resulting in twelve distinct spots. The presence of either one or two epitaxial growth relations in the CCSS film depends on the epitaxial growth of the Pt buffer on $Al_2O_3$(0001), which can follow either OR1 or OR2.[35,36] For the CCSS films in Figure 3b and 3c, their respective Pt buffers grow with OR1 relation and a combination of OR1 and OR2 relations (see Supporting Information Figure S3). Accordingly, the CCSS films adopt the growth relations of their Pt buffer. In essence, the Pt buffer induces epitaxial growth of the CCSS film on $Al_2O_3$(0001). This growth process is robust across a broad composition range and works equally well for Ru-free and Ru-rich CCSS composition (see Supporting Information Figure S4).

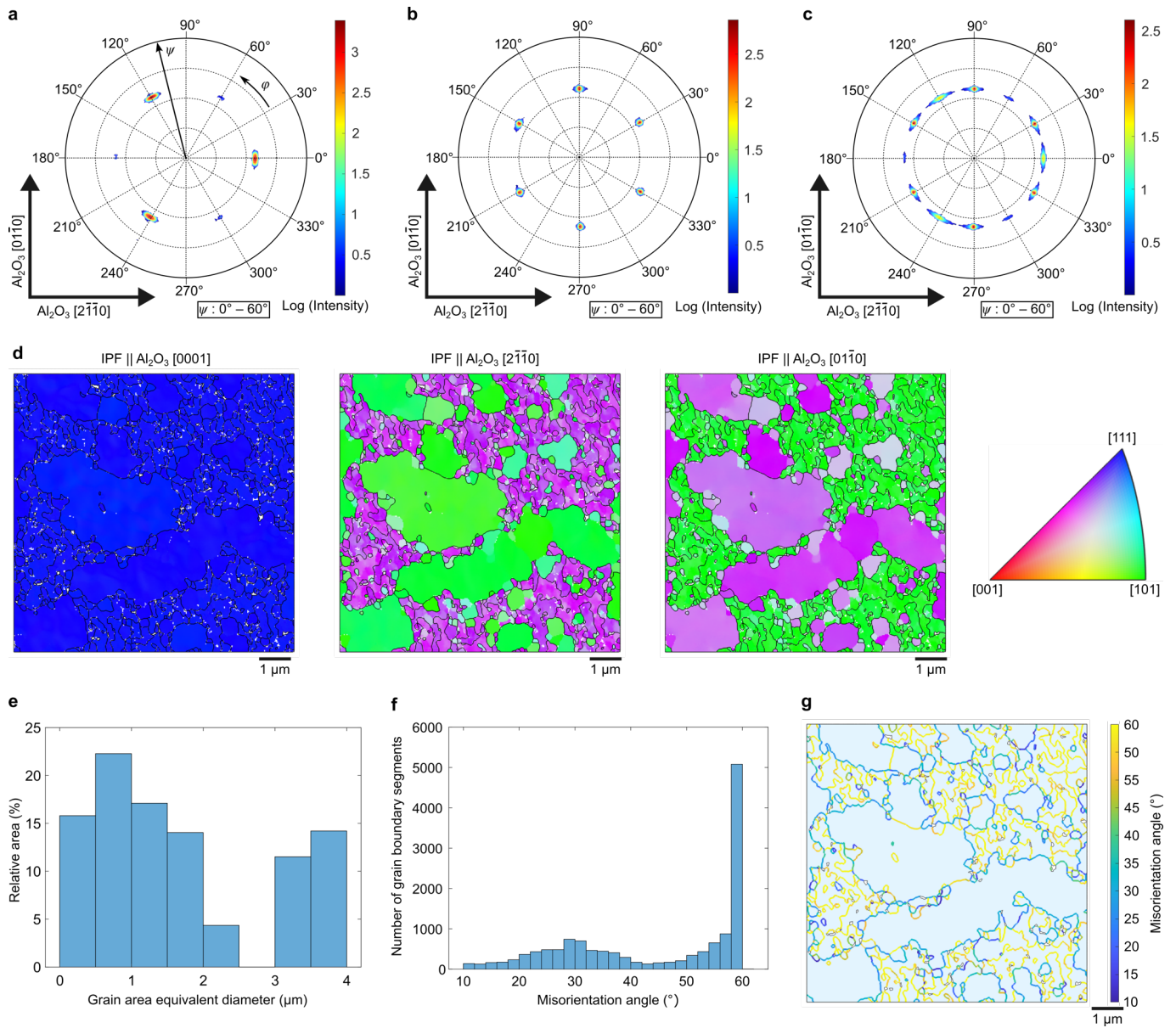

**Figure 3.** *Epitaxial growth and microstructure analyses of CCSS films.* {220} pole figures of (a) 60 nm Pt film, (b) 60 nm CCSS film grown on 5 nm Pt buffer and (c) 100 nm CCSS film grown on 40 nm Pt buffer. As described in the text, epitaxial growth is observed in all cases with {111} twinning. Two epitaxial growth relations, OR1 and OR2, are observed in (c), while only OR2 is observed in (a) and OR1 in (b). The extended pole figures covering $\psi$ from 0° to 80° are included in Supporting Information Figure S5. (d) Inverse pole figure (IPF) maps of a typical microstructure in the 100 nm CCSS film along $Al_2O_3$ [0001], [2$\bar{1}\bar{1}$0], and [01$\bar{1}$0] crystallographic directions. Grain boundaries are highlighted in black, the crystal orientation is color-coded, and the unindexed pixels appear in white. The regions with OR1 and OR2 relations appear as purple and green, respectively, in the IPF map along $Al_2O_3$[2$\bar{1}\bar{1}$0]. Plots of (e) grain size distribution and (f) grain boundary misorientation distribution in the investigated region. Grain boundaries with 30° misorientation separate OR1 and OR2 grains. Grain boundaries with 60° misorientation correspond to {112}Σ3 boundaries. (g) Areal distribution map of grain boundaries colored by their misorientation. The {112}Σ3 boundaries are primarily located in OR1 regions.

To examine microstructural characteristics associated with the two epitaxial relations, we performed EBSD of the 100 nm CCSS film. Figure 3d shows inverse pole figure (IPF) maps of the film's typical microstructure along the out-of-plane direction $Al_2O_3[0001]$ and the two orthogonal in-plane directions: $Al_2O_3[2\bar{1}\bar{1}0]$ and $Al_2O_3[01\bar{1}0]$. The examined region is decomposed into grains using a misorientation threshold of 10°, with grain boundaries marked in black and crystal orientation indexed as per the included colormap. The grains display a [111] type direction normal to the film surface, while two distinct in-plane directions ([101] type in green and [112] type in purple), align with the $[2\bar{1}\bar{1}0]$ direction of the $Al_2O_3$ substrate. Indeed, these two colors in the IPF map along $Al_2O_3[2\bar{1}\bar{1}0]$ correspond to the epitaxial growth relations OR2 and OR1 observed in the pole figure in Figure 3c. The IPF maps reveal that the larger grains are predominantly associated with the OR2 relation. The grain size distribution presented in Figure 3e affirms that the majority of the grains are micrometer-sized, with some reaching up to 4 μm. The distribution of misorientation at the grain boundaries (Figure 3f) peaks at 30° and 60°. The peak at 30° corresponds to the grain boundaries between OR2 and OR1 grains, and the peak at 60° corresponds to {112}Σ3 twin boundaries originating from twinning in the CCSS film during growth. The areal correlation map (Figure 3g) shows that {112}Σ3 boundaries are located densely in OR1 regions. These microstructural differences between OR2 and OR1 regions have been observed previously in epitaxial Al(111) films grown on $Al_2O_3$(0001) and were attributed to a higher interfacial energy and mobility of {112}Σ3 boundaries in OR2 relation than in OR1 relation.[41–43] Accordingly, OR2 regions tend to form larger grains with fewer Σ3 boundaries, which is also consistent with our observations in the epitaxial CCSS film.

To investigate compositional homogeneity and cross-sectional morphology of the 100 nm epitaxial CCSS film, we performed (S)TEM of a cross-sectional lamella prepared by focused ion beam milling (see Methods). In the overview high-angle annular dark field (HAADF)-STEM image (Figure 4a), the CCSS film and the Pt buffer are clearly distinguishable from the Z-contrast. STEM-EDS elemental maps performed inside the CCSS film (Figure 4b) reveal a uniform distribution of the constituent elements, with $Ir_{21}Pd_{21}Pt_{19}Rh_{17}Ru_{22}$ as the film composition. Despite the compositional homogeneity, several defects are observed in the CCSS film and the Pt buffer (Figure 4c, d, f, g). We observe dislocations (Figure 4c) and {112}Σ3 boundaries (Figure 4d), which continue across the buffer–CCSS film interface to the film surface. Dislocations commonly form in heteroepitaxial films due to lattice mismatch between film and substrate.[44–46] Selected area diffraction (SAD) pattern analysis of a film region indicated by red circles in the bright-field (BF) and dark-field (DF) TEM images (Figure 4d) confirms the occurrence of twinning with two sets of spots marked in dark and light orange. Furthermore, spots at $1/3g_{111}$ are visible in the SAD, suggesting the occurrence of the 9R structure. Using such a spot (encircled in green), the corresponding DF TEM image highlights the areal distribution of the 9R structure, appearing bright in the DF image. The 9R regions are observed in both the CCSS film and Pt buffer, which in some cases extend from the substrate-buffer interface to the CCSS film surface. On examining the cross-section lamella by high-resolution TEM, the Pt buffer–CCSS film interface appears coherent (Figure 4e), as the CCSS film continues to grow epitaxially on the Pt buffer. The high-resolution imaging of a potentially 9R containing region in the CCSS film (Figure 4f) exhibits contrast as that of periodically spaced stacking faults, which begin about 2 nm above the interface with the buffer layer. The FFT patterns of locations, highlighted by blue and red dotted squares in the image, reveal an FCC structure with $[1\bar{1}0]$ zone-axis in the Pt buffer and a 9R structure in the CCSS film with additional spots in the diffraction pattern appearing at every $1/3g_{111}$ in $[1\bar{1}0]$ zone-axis of CCSS (Figure 4f). A magnified view of the 9R defect region is shown in Figure 4g, where it is bound by a coherent (111) plane (solid green line) and an incoherent boundary (dotted green line). The 9R structure is characterized by a periodically occurring stacking fault every third {111} plane in the FCC stacking sequence. This defect forms due to dissociation of {112}Σ3 incoherent twin boundary and has been previously reported in FCC metals with low intrinsic stacking fault energy, such as Ag,[47] Au,[48] Cu,[49,50] Pd,[51], and even in some CCSS systems.[52,53] While Ir, Pt, and Rh possess high intrinsic stacking fault energy, the presence of Pd and Ru may lower the intrinsic stacking fault energy of the investigated CCSS.[54,55] Our observations of 9R structure in both the Pt buffer and the CCSS film suggest that their occurrence is likely linked to the non-equilibrium nature of film growth and low intrinsic stacking fault energy of CCSS. Overall, these results show that epitaxial films can also serve as a model platform for investigating complex defect structures of CCSS systems.

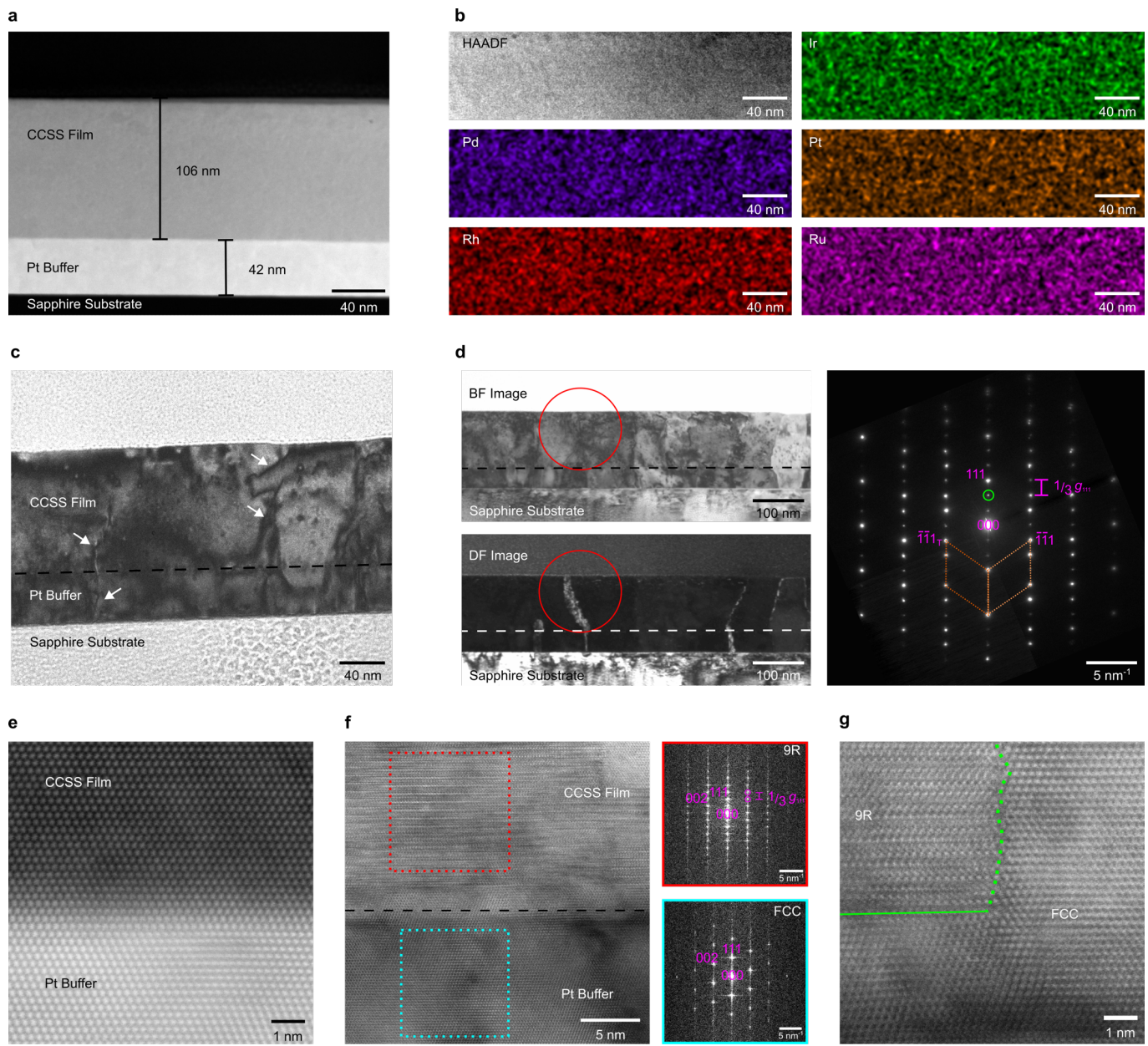

**Figure 4.** *Compositional homogeneity and defect analyses of an epitaxial CCSS film*. (a) HAADF-STEM image depicting the cross-sectional overview of the 100 nm CCSS film grown on 40 nm Pt buffer with Al$_2$O$_3$(0001) substrate. (b) STEM-EDS elemental maps of a region within the CCSS film, depicted in the HAADF image, demonstrate homogenous distribution of the elements. (c) Two-beam BF TEM image of the film cross-section revealing dislocations, indicated by arrows. (d) BF TEM image of the film cross-section in [1$\bar{1}$0] zone axis; red circle highlights the region used for selected area diffraction (SAD). The diffraction pattern exhibits two sets of spots (indicated by dark and light orange) originating from the {112}Σ3 twin relation. Additional low intensity spots appear at 1/3$g_{111}$, suggesting the presence of 9R structure. DF TEM image obtained with a 1/3$g_{111}$ type reflection highlighted by green circle shows regions with 9R structure in the cross-section. (e) HAADF-STEM image of the buffer-film interface, showing coherent arrangement of atoms, i.e., epitaxial growth. (f) BF STEM image depicting a region with finely spaced periodic stacking faults (indicated by red dotted square) in the CCSS film. FFT analyses of regions indicated by blue and red dotted squares reveal FCC structure in [1$\bar{1}$0] zone-axis in the Pt buffer and 9R structure in the defect region of CCSS film in [1$\bar{1}$0] zone-axis. (g) BF STEM image illustrating a 9R region separated by an incoherent boundary (green dotted line) and a coherent (111) plane (green line) from FCC region in the CCSS film.

The primary purpose of the epitaxial CCSS platform is to enable high resolution studies on their electrocatalytic properties. The 100 nm epitaxial CCSS film with its microstructural heterogeneities is well-suited to showcase correlative electrochemical characterization, which enables understanding the influence of microstructure on electrocatalytic properties. To perform the correlative study involving multiple techniques, we developed a navigation approach using micro-indents on the film as markers (see Supporting Information Section 1). Figure 5a depicts the EBSD IPF map of a region of interest (ROI) enclosing both the OR1(purple) and the OR2 (green) related grains. Its corresponding surface topography (Figure 5b) reveals that the RMS roughness of the OR2 region (0.53 nm) is slightly lower than that of the OR1 region (0.75 nm) (see Supporting Information Figure S6 for the examined locations). To evaluate the electrocatalytic activity of OR1 and OR2 grains for the hydrogen evolution reaction (HER), we conducted localized electrochemical measurements at the nanoscale by SECCM. A single-barrel nanopipette with an opening diameter of approximately 85 nm (see Supporting Information Figure S7), filled with 0.1 M $HClO_4$ electrolyte, was used for the experiment. The HER activity in the ROI was spatially resolved by recording a series of cyclic voltammograms over a rectangular grid of 47 × 34 measurement areas, using a hopping distance of 150 nm (see Methods for set-up and measurement protocol). As illustrated in Figure 5c, the SECCM map reveals two distinct regions with higher and lower current values at -0.4 V vs. RHE. By correlating with the IPF map of the ROI, OR2 and OR1 regions can be correlated with the different activity regions visualized in the SECCM map (Figure 5d), thereby allowing us to analyze the current distribution grouped by the epitaxial relation (Figure 5e). From our statistically robust examination of the CCSS film surface, it is evident that the OR1 region exhibits higher HER activity (mean current: −0.284 ± 0.003 nA) than the OR2 region (mean current: −0.219 ± 0.002 nA). A video illustrating the evolution of current with applied potential in the ROI is available in Supporting Information Video S1 and the averaged voltammograms categorized by their epitaxial relation are shown in Figure 5f.

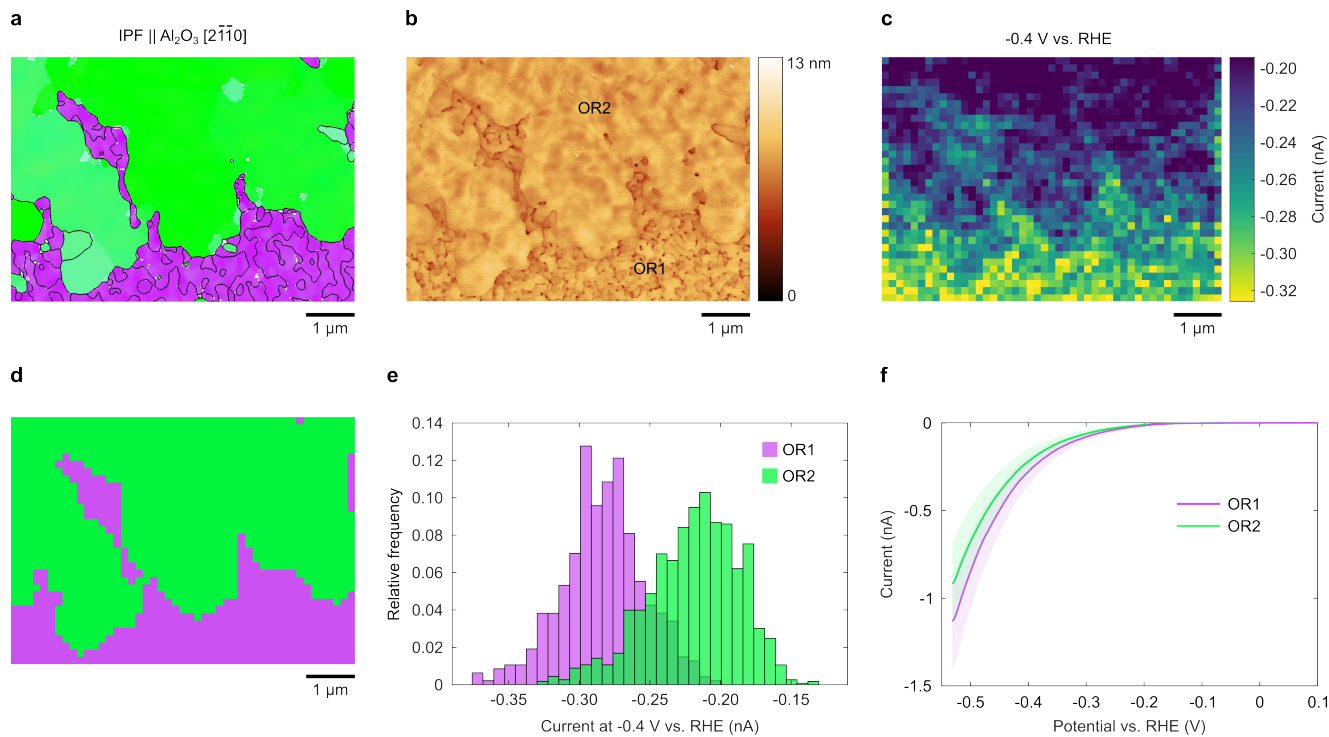

**Figure 5.** *Correlative epitaxial orientation-electrocatalysis investigations on CCSS film.* (a) Distortion-corrected EBSD IPF map along $Al_2O_3[2\bar{1}\bar{1}0]$ of a ROI in the 100 nm CCSS film containing both OR1 and OR2 grains. (b) Surface topography of the ROI. (c) SECCM current map of the ROI depicting the HER activity recorded at -0.4 V vs. RHE in 0.1 M $HClO_4$ electrolyte. (d) Classification of the SECCM map into OR1 and OR2 regions by correlation with the IPF map. (e) Distribution of the recorded current at -0.4 V vs. RHE grouped by the epitaxial growth relation. (f) Averaged voltammograms of the OR1 and OR2 regions within the SECCM map using the measured areas classified in (d). The shadow regions indicate the 95% confidence interval.

To rule out that the differences in HER activity between OR1 and OR2 regions originate from differences in surface composition, we performed Auger electron spectroscopy on OR1 and OR2 regions. The results show similar composition in both regions (see Supporting Information Figure S8). From the topography scan of the ROI (Figure 5b), we previously noted that the OR1 regions exhibit slightly higher RMS roughness than the OR2 regions. The topographic variations within the OR1 region are particularly localized at the grain boundaries. EBSD analysis of these boundaries (Supporting Information Figure S9) reveal them as {112}Σ3 incoherent twin boundaries. We propose that these incoherent twin boundaries provide additional active sites for HER, thereby contributing to higher HER activity in the OR1 region than the OR2 region.[56–58]

## 3. Summary

In summary, our results demonstrate the potential and importance of correlative scale-bridging investigations in CCSS systems, which is achievable with epitaxial films and navigation structures. We have demonstrated a reliable approach for achieving epitaxial growth of CCSS films with micrometer-sized grains and smooth surfaces. This advancement, together with a facile approach for contamination-free surface navigation, provide a model platform for scale-bridging and correlative structure–activity investigation of CCSSs at the nanoscale. The epitaxial growth approach works over a broad composition range, which enables fundamental investigations into promising CCSS compositions identified by high-throughput exploration. Hence, we anticipate epitaxial films to significantly aid in accelerating the mechanistic understanding and rational designing of CCSS electrocatalysts.

4. Methods

*Thin film growth*

Thin films of the CCSS system Ir-Pd-Pt-Rh-Ru were fabricated by co-sputtering 50.8 mm diameter elemental targets in an 8-cathode sputter system (AJA International ATC-2200-UHV, base pressure: $< 2.6 \times 10^{-6}$ Pa). The films were deposited either on single-side polished *c*-plane (0001) sapphire wafers of 100 mm diameter or on 10 mm × 10 mm diced pieces of the wafer for better fit with various characterization techniques. DC and RF power sources were used for the depositions and assigned as per the sputter yield of the element. The Ar (6N) process pressure was 0.67 Pa. The sample stage rotated at 20 RPM during deposition to achieve uniform film composition. To achieve epitaxial growth, some depositions were performed on heated substrates and using a Pt buffer layer. The deposition parameters of each sample mentioned in the main text are summarized in Supporting Information Table S1.

*Structure and microstructure characterization*

The crystal structure of the CCSS films grown directly on $Al_2O_3(0001)$ substrate was characterized using an X-ray diffractometer (Bruker D8 Discover) with Cu $K_\alpha$ X-ray radiation ($\lambda = 0.15418$ nm) and a 2D detector (VANTEC 500) in coupled $\theta$-$2\theta$ mode. Three overlapping diffractogram frames were recorded during a measurement at 30°, 55°, and 80°. These three frames were then stitched and integrated to obtain a 1-D diffractogram using Bruker DIFFRAC.EVA software. The crystal structure of epitaxial Pt film sample and epitaxial CCSS films grown with a Pt buffer was characterized using a X-ray diffractometer (Panalytical X'Pert) equipped with an Eulerian cradle, x-y-z sample stage, Cu $K_\alpha$ X-ray tube with point focus ($\lambda = 0.15418$ nm), polycapillary X-ray lens with adjustable cross slits on the primary beam side, parallel slits collimator on the diffracted beam side, and a PIXcel1D detector used in 0D mode. The coupled $\theta$-$2\theta$ scans were performed in the $2\theta$ range of 35° to 95°, with 0.005° steps. This instrument setup was also used to perform {220} pole figure measurements and $\varphi$ scans to confirm epitaxial growth of the films and determine the epitaxial growth relationships. For the pole figure measurements, the sample tilt ($\psi$) was varied from 0° to 80° and the sample rotation ($\varphi$) was varied from 0° to 360°, with 1° step size.

The surface morphology and roughness of the fabricated films (Figure 2) was investigated using an AFM (Bruker FastScan) in PeakForce tapping mode with a ScanAsyst-Air tip. The obtained raw data were processed using Gwyddion software. The surface morphology of the region investigated for electrochemical activity (Figure 5b) was characterized by an AFM (Park Systems NX10) in non-Contact Mode using a PPP-NCHR tip.

EBSD of the 100 nm epitaxial CCSS film was performed using a scanning electron microscope (JEOL JSM-7200F) fitted with an EBSD detector (Oxford Instruments AZtec Symmetry). The EBSD scans were performed using standard EBSD operating parameters with a step size of 30 nm. The EBSD data were analyzed using MTEX toolbox in MATLAB. The EBSD map used for correlative investigation (Figure 5a) was corrected for distortion using the AFM map of the corresponding region as reference with a custom MATLAB script. The volume composition of the CCSS films without Pt buffer layer was determined in this SEM by EDS using Oxford Instruments Oxford Aztec Energy X-MaxN detector (accuracy: 1 at.%). A FIB system (FEI Helios G5 CX) was used to prepare TEM lamella of 100 nm CCSS film cross-section. A probe-corrected TEM (Titan Themis 300 equipped with Super-X EDS detector) was used for STEM imaging and STEM-EDS mapping of the lamella at 300 kV. The two-beam TEM imaging of lamella was done in a TEM (JEOL JEM-2100Plus) operated at 200 kV. SAD acquisition, BF and DF imaging were performed in a TEM (Tecnai F20 G$^2$) operated at 200 kV. Auger electron spectroscopy of the 100 nm epitaxial CCSS film was conducted in a Scanning Auger Nanoprobe 710, operated at 15 kV. The sample surface was sputter cleaned using Ar$^+$ ions at 1 kV for an optimum duration to get rid of carbon contamination with minimal impact on the actual film surface. Micro-indents on the 100 nm CCSS film as navigation markers were made using a microhardness tester (KB 30SR) according to DIN EN ISO 6507 standard using a force of 2.94 N.

*Electrochemical Characterization*

The electrochemical experiments were conducted using a custom-built SECCM setup. In this configuration, a single-barrel SECCM pipette filled with 0.1 M HClO$_4$ electrolyte was mounted on a stationary holder, and the film sample was placed inside a chamber connected to the positioning system. The chamber is constructed from acrylonitrile–butadiene–styrene (ABS) polymer wrapped in Al foil and continuously purged with 198 mL min$^{-1}$ of humidified Ar to ensure an inert atmosphere and prevent interference from O$_2$. The Ar flow is controlled by a mass flow controller (EL-Flow Select F-201CV-100, Bronkhorst Instruments GmbH). A coarse positioning of the sample (about 40 μm below pipette and about 100 μm from an indent marker) was performed using a motorized precision x–y–z stages (Owis) controlled by an LStep PCIe controller (Lang), assisted by an optical imaging system, consisting of a VZM™ 200i zoom lens (Edmund Optics) and a Motic camera. Fine positioning during the SECCM scan was performed using an x–y–z piezo positioner (P-611.3S NanoCube, Physik Instrumente) mounted beneath the sample holder. The entire setup is mounted on a benchtop passive vibration isolation platform (50BM-8, Minus K Technology) placed on a vibration isolation workstation (M-VIS3648-SG2-325A, Newport) to minimize mechanical interference. The platform is surrounded by thermal insulation panels (VakuVIP B2, Vaku-Isotherm GmbH) and a copper mesh Faraday cage (VIS-FDC-3648, Newport) to ensure low electrical noise and thermally stable environment during the experiment.

To perform the electrochemical experiments, a potential was applied to the working electrode relative to the quasi-reference counter-electrode (QRCE), and the resulting current was measured at the QRCE using a low-noise potentiostat (VA-10x, npi electronics). Data acquisition and instrument control were performed using an FPGA card (USB-7855R) and a modified LabVIEW (National Instruments) program based on the Warwick Electrochemical Scanning Probe Microscopy (WEC-SPM) toolbox. The data were acquired at 488 Hz. Single-barrel pipettes were fabricated by pulling single-barrel quartz glass capillaries (QF120-90-10, Sutter Instruments) using a $CO_2$-laser puller (P-2000, Sutter Instruments). The quartz capillaries were cleaned with wipes soaked in ethanol. Mapping was performed in 'scan-hopping mode', where the SECCM probe was moved sequentially across a predefined grid of 47 × 34 points, equally spaced by 150 nm, covering an area of approximately 4.95 × 6.90 µm$^2$. At each position, the pipette approached the surface, polarized at –0.45 V vs. Ag/AgCl (3.4 M KCl), with an approach rate of 0.7 µm s$^{-1}$. The approach was halted upon detection of the current signal, signifying contact between the electrolyte meniscus and the CCSS film surface. This established a two-electrode electrochemical cell: the CCSS film as the working electrode, and the leak-free Ag/AgCl (3.4 M KCl) electrode serving as both counter and quasi-reference electrode. After the approach, in each point, a cyclic voltammetry scan between -0.160 V and -0.810 V vs. Ag/AgCl (3.4 M KCl) was performed at 2 V s$^{-1}$ to evaluate the electroactivity for HER. The results are reported in the reversible hydrogen electrode (RHE) scale. Accordingly, the conversion of the recorded potential was done following the equation:

$$E_{RHE} \text{ [V]} = E_{Ag/AgCl(3.4M\ KCl)} + 0.210 + E_{OCP} + 0.0592 \times pH$$

where $E_{Ag/AgCl(3.4\ M\ KCl)}$ is the applied potential versus the quasi-reference counter electrode (QRCE), 0.210 V is the standard potential of the Ag/AgCl (3 M KCl) reference electrode at 25 °C, $E_{OCP}$ is the measured open circuit potential difference between the QRCE and the Ag/AgCl (3 M KCl) reference electrode. Data processing, visualization, and analysis were carried out using MATLAB. The average current values at -0.40 V vs. RHE grouped by the two regions with distinct epitaxial relationships are presented in the manuscript with their 95 % confidence intervals and were calculated as:

$$mean \pm z_{\alpha/2} \times S.E.$$

where $z_{\alpha/2}$ is the critical value from the standard normal distribution and $S.E.$ is the standard error.

**Data availability**

Data supporting the findings of this study are openly available at Zenodo DOI:10.5281/zenodo.18745470.

# References

1. Zhang, Y. *et al.* Microstructures and properties of high-entropy alloys. *Prog. Mater. Sci.* **61**, 1–93 (2014).

2. Ye, Y. F., Wang, Q., Lu, J., Liu, C. T. & Yang, Y. High-entropy alloy: challenges and prospects. *Mater. Today.* **19**, 349–362 (2016).

3. Pickering, E. J. & Jones, N. G. High-entropy alloys: a critical assessment of their founding principles and future prospects. *Inter. Mater. Rev.* **61**, 183–202 (2016).

4. Han, L. *et al.* Multifunctional high-entropy materials. *Nat. Rev. Mater.* **9**, 846–865 (2024).

5. Löffler, T. *et al.* Discovery of a multinary noble metal–free oxygen reduction catalyst. *Adv. Energy Mater.* **8**, 1802269 (2018).

6. Luo, H. *et al.* A strong and ductile medium-entropy alloy resists hydrogen embrittlement and corrosion. *Nat. Commun.* **11**, 3081 (2020).

7. Xin, Y. *et al.* High-entropy alloys as a platform for catalysis: progress, challenges, and opportunities. *ACS Catal.* **10**, 11280–11306 (2020).

8. Sun, Y. & Dai, S. High-entropy materials for catalysis: A new frontier. *Sci. Adv.* **7**, eabg1600 (2021).

9. Han, L. *et al.* Strong and ductile high temperature soft magnets through Widmanstätten precipitates. *Nat. Commun.* **14**, 8176 (2023).

10. Li, Z. *et al.* Strength, plasticity and coercivity tradeoff in soft magnetic high-entropy alloys by multiple coherent interfaces. *Acta Mater.* **254**, 118970 (2023).

11. Löffler, T. *et al.* Toward a paradigm shift in electrocatalysis using complex solid solution nanoparticles. *ACS Energy Lett.* **4**, 1206–1214 (2019).

12. Batchelor, T. A. A. *et al.* High-entropy alloys as a discovery platform for electrocatalysis. *Joule* **3**, 834–845 (2019).

13. Löffler, T., Ludwig, A., Rossmeisl, J. & Schuhmann, W. What makes high-entropy alloys exceptional electrocatalysts? *Angew. Chem. Int. Ed.* **60**, 26894–26903 (2021).

14. Yao, Y. *et al.* Carbothermal shock synthesis of high-entropy-alloy nanoparticles. *Science* **359**, 1489–1494 (2018).

15. Zhang, C. *et al.* Rapid synthesis of subnanoscale high-entropy alloys with ultrahigh durability. *Nat. Mater.* **25,** 26–34 (2026).

16. Ludwig, A. Discovery of new materials using combinatorial synthesis and high-throughput characterization of thin-film materials libraries combined with computational methods. *npj Comput. Mater.* **5**, 70 (2019).

17. Banko, L. *et al.* Unravelling composition–activity–stability trends in high entropy alloy electrocatalysts by using a data-guided combinatorial synthesis strategy and computational modeling. *Adv. Energy Mater.* **12**, 2103312 (2022).

18. Tetteh, E. B. *et al.* Long-Range SECCM enables high-throughput electrochemical screening of high entropy alloy electrocatalysts at up-to-industrial current densities. *Small. Methods.* **8**, 2301284 (2024).

19. Thelen, F. *et al.* Accelerating combinatorial electrocatalyst discovery with bayesian optimization: a case study in the quaternary System Ni-Pd-Pt-Ru for the oxygen evolution reaction. *Adv. Sci.* **12**, e07302 (2025).

20. Clausen, C. M., Nielsen, M. L. S., Pedersen, J. K. & Rossmeisl, J. Ab initio to activity: machine learning-assisted optimization of high-entropy alloy catalytic activity. *High-Entropy Alloy. Mater.* **1**, 120–133 (2023).

21. Batchelor, T. A. A. *et al.* Complex-solid-solution electrocatalyst discovery by computational prediction and high-throughput experimentation. *Angew. Chem. Int. Ed.* **60**, 6932–6937 (2021).

22. Wang, C. *et al.* Facet-Controlled Synthesis of platinum-group-metal quaternary Alloys: the case of nanocubes and {100} facets. *J. Am. Chem. Soc.* **145**, 2553–2560 (2023).

23. Wu, C.-Y. *et al.* A catalyst family of high-entropy alloy atomic layers with square atomic arrangements comprising iron- and platinum-group metals. *Sci. Adv.* **10**, eadl3693 (2024).

24. Hsiao, Y.-C. *et al.* A library of seed@high-entropy-alloy core–shell nanocrystals with controlled facets for Catalysis. *Adv. Mater.* **37**, 2411464 (2025).

25. Wang, Y., Li, M., Gordon, E., Ye, Z. & Ren, H. Nanoscale colocalized electrochemical and structural mapping of metal dissolution reaction. *Anal. Chem.* **94**, 9058–9064 (2022).

26. Wahab, O. J. *et al.* Proton transport through nanoscale corrugations in two-dimensional crystals. *Nature* **620**, 782–786 (2023).

27. Martín-Yerga, D., Unwin, P. R., Valavanis, D. & Xu, X. Correlative co-located electrochemical multi-microscopy. *Curr. Opin. Electrochem.* **42**, 101405 (2023).

28. Xu, X. *et al.* Surface structure and grain boundary effects on the oxygen evolution reaction at gold electrodes. *ACS Electrochem.* **1**, 1852–1862 (2025).

29. Pfisterer, J. H. K., Liang, Y., Schneider, O. & Bandarenka, A. S. Direct instrumental identification of catalytically active surface sites. *Nature* **549**, 74–77 (2017).

30. Addab, Y. *et al.* Microstructure evolution and thermal stability of equiatomic CoCrFeNi films on (0001) α-$Al_2O_3$. *Acta Mater.* **200**, 908–921 (2020).

31. Kini, M. K. *et al.* Equiatomic CoCrFeNi thin films on c-sapphire: the role of twins and orientation relationships. *Adv. Eng. Mater.* **26**, 2400720 (2024).

32. Schwarz, H. *et al.* Fabrication of single-crystalline CoCrFeNi thin films by DC magnetron sputtering: a route to surface studies of high-entropy alloys. *Adv. Mater.* **35**, 2301526 (2023).

33. Ledieu, J. *et al.* Epitaxial growth of rare-earth high-entropy alloy thin films. *ACS Nano* **19**, 26400–26410 (2025).

34. Wu, D. *et al.* On the electronic structure and hydrogen evolution reaction activity of platinum group metal-based high-entropy-alloy nanoparticles. *Chem. Sci.* **11**, 12731–12736 (2020).

35. Vargas, R., Goto, T., Zhang, W. & Hirai, T. Epitaxial growth of iridium and platinum films on sapphire by metalorganic chemical vapor deposition. *Appl. Phys. Lett.* **65**, 1094–1096 (1994).

36. Torres-Castanedo, C. G. *et al.* Ultrasmooth epitaxial Pt thin films grown by pulsed laser deposition. *ACS Appl. Mater. Interfaces.* **16**, 1921–1929 (2024).

37. Aleman, A. *et al.* Ultrahigh vacuum dc magnetron sputter-deposition of epitaxial Pd(111)/$Al_2O_3$(0001) thin films. *J. Vac. Sci. Technol. A.* **36**, 030602 (2018).

38. Majer, L. N. *et al.* Growth of high-quality ruthenium films on sapphire. *J. Vac. Sci. Technol. A.* **42**, 052702 (2024).

39. Wagner, T., Richter, G. & Rühle, M. Epitaxy of Pd thin films on (100) SrTiO$_3$: A three-step growth process. *J. Appl. Phys.* **89**, 2606–2612 (2001).

40. Hildner, M. L., Minvielle, T. J. & Wilson, R. J. Epitaxial growth of ultrathin Pt films on basal-plane sapphire: the emergence of a continuous atomically flat film. *Surf. Sci.* **396**, 16–23 (1998).

41. Hieke, S. W., Breitbach, B., Dehm, G. & Scheu, C. Microstructural evolution and solid state dewetting of epitaxial Al thin films on sapphire (α-Al$_2$O$_3$). *Acta Mater.* **133**, 356–366 (2017).

42. Saood, S., Brink, T., Liebscher, C. H. & Dehm, G. Influence of variation in grain boundary parameters on the evolution of atomic structure and properties of [111] tilt boundaries in aluminum. *Acta Mater.* **268**, 119732 (2024).

43. Barmak, K. *et al.* Grain boundary energy and grain growth in Al films: Comparison of experiments and simulations. *Scr. Mater.* **54**, 1059–1063 (2006).

44. Hull, R. & Bean, J. C. Misfit dislocations in lattice-mismatched epitaxial films. *Crit. Rev. Solid State Mater. Sci.* **17**, 507–546 (1992).

45. Bennett, S. E. Dislocations and their reduction in GaN. *Mater. Sci. and Technol.* **26**, 1017–1028 (2010).

46. Kar, S. *et al.* Growth twins and premartensite microstructure in epitaxial Ni-Mn-Ga films. *Acta Mater.* **252**, 118902 (2023).

47. Ernst, F. *et al.* Theoretical prediction and direct observation of the 9R structure in Ag. *Phys. Rev. Lett.* **69**, 620–623 (1992).

48. Medlin, D. L., Foiles, S. M. & Cohen, D. A dislocation-based description of grain boundary dissociation: application to a 90° ⟨110⟩ tilt boundary in gold. *Acta Mater.* **49**, 3689–3697 (2001).

49. Medlin, D. L., Campbell, G. H. & Carter, C. B. Stacking defects in the 9R phase at an incoherent twin boundary in copper. *Acta Mater.* **46**, 5135–5142 (1998).

50. Wang, J., Anderoglu, O., Hirth, J. P., Misra, A. & Zhang, X. Dislocation structures of Σ3 {112} twin boundaries in face centered cubic metals. *Appl. Phys. Lett.* **95**, 021908 (2009).

51. Amin-Ahmadi, B. *et al.* High resolution transmission electron microscopy characterization of fcc → 9R transformation in nanocrystalline palladium films due to hydriding. *Appl. Phys. Lett.* **102**, 071911 (2013).

52. Lu, W. *et al.* Interfacial nanophases stabilize nanotwins in high-entropy alloys. *Acta Mater.* **185**, 218–232 (2020).

53. Liu, X. *et al.* Multi-principal element alloys with high-density nanotwinned 9R phase. *Mater. Des.* **229**, 111925 (2023).

54. Rosengaard, N. M. & Skriver, H. L. Calculated stacking-fault energies of elemental metals. *Phys. Rev. B* **47**, 12865–12873 (1993).

55. Cai, J., Wang, F., Lu, C. & Wang, Y. Y. Structure and stacking-fault energy in metals Al, Pd, Pt, Ir, and Rh. *Phys. Rev. B* **69**, 224104 (2004).

56. Liang, Y., Csoklich, C., McLaughlin, D., Schneider, O. & Bandarenka, A. S. Revealing active sites for hydrogen evolution at Pt and Pd atomic Layers on Au surfaces. *ACS Appl. Mater. Interfaces.* **11**, 12476–12480 (2019).

57. Santana, J. A., Cruz, B., Melendez-Rivera, J. & Rösch, N. Strain and low-coordination effects on monolayer nanoislands of Pd and Pt on Au(111): a comparative analysis based on density functional results. *J. Phys. Chem. C* **124**, 13225–13230 (2020).

58. Schmidt, T. O. *et al.* Elucidation of structure–activity relations in proton electroreduction at Pd surfaces: theoretical and experimental study. *Small* **18**, 2202410 (2022).

**Author Contributions**

S.K. led the conceptualization and experimental design for film growth and marker-based navigation; performed sample fabrication, XRD, AFM, and EBSD characterization, formal analysis, and wrote the first manuscript draft. A.E-M. designed marker-based navigation experiments, performed SECCM and AFM measurements, as well as formal analysis. H.X. and M.J. performed TEM characterization and formal analysis supported by C. Somsen and C. Scheu. K.Y. performed XRD pole figure measurements. J.P-M. performed the micro-indentation marking. A.L. contributed to the conceptualization and writing of the manuscript. C. Scheu., C.A., and A.L. provided supervision, project administration, and resources. All authors reviewed and contributed to the final version of the manuscript.

**Acknowledgements**

This research was financially supported by the Deutsche Forschungsgemeinschaft (DFG, German Research Foundation) CRC 1625, project number 506711657. S.K. and A.L. acknowledge support through subproject A02. A.M. and C.A. acknowledge support through subproject C02. H.X., M.J., C. Somsen, and C. Scheu acknowledge support through subproject B03. J.P-M. acknowledges support through subproject S. The authors thank Dr. Aleksander Kostka (subproject S) for the preparation of TEM lamella. The authors acknowledge ZGH, Ruhr University Bochum for AFM, EBSD, EDS, FIB, micro-indentation, and XRD of the films.

**Competing Interests**

The authors declare no competing interests.

**Supporting Information**

**Table S1.** Overview of sputter deposition parameters used for film growth. Power values indicated with an asterisk (*) represent RF power; other values correspond to DC power. The Pt buffer was deposited at 40 W DC (Fig. 2f, Fig. S4a, d) or 60 W RF power (Fig. 2g).

| Reference in Manuscript | Nominal film thickness (nm) | Pt buffer Thickness (nm) | Deposition Temperature (K) | Sputter Power (W) | | | | |
|---|---|---|---|---|---|---|---|---|
| | | | | Ir | Pd | Pt | Rh | Ru |
| Fig. 2a | 60 | - | 297 | 24 | 33* | 11 | 59* | 43 |
| Fig. 2b | 60 | - | 873 | 22 | 32* | 10 | 57* | 36 |
| Fig. 2c | 60 | - | 673 (45 nm) 873 (15 nm) | 34 | 12 | 36* | 65* | 50 |
| Fig. 2e | 60 | - | 873 | - | - | 40 | - | - |
| Fig. 2f | 60 | 5 | 873 | 24 | 33* | 11 | 59* | 43 |
| Fig. 2g | 100 | 40 | 873 | 19 | 30* | 32* | 49* | 28 |
| Fig. S4a | 100 | 40 | 873 | 13 | 21* | 11 | 31* | 65 |
| Fig. S4d | 100 | 40 | 873 | 43 | 38* | 15 | 63* | - |
| - | 60 | - | 296 | 13 | 21* | 11 | 31* | 65 |

**Table S2.** Overview of the EDS film compositions of the investigated CCSS films. Values marked with an asterisk (*) belong to CCSS films produced with identical sputter power conditions at room temperature without a Pt buffer.

| Reference in Manuscript | EDS Film composition |
|---|---|
| Figure 2a | $Ir_{20}Pd_{20}Pt_{17}Rh_{20}Ru_{23}$ |
| Figure 2b | $Ir_{23}Pd_{16}Pt_{16}Rh_{22}Ru_{23}$ |
| Figure 2c | $Ir_{20}Pd_{19}Pt_{17}Rh_{20}Ru_{24}$ |
| Figure 2f, 3b, S1a, S3a, S5b | $Ir_{20}Pd_{20}Pt_{17}Rh_{20}Ru_{23}$* |
| Figure 2g, 3c, 3d-g, 4, 5, S1b, S3b, S5c, S6, S8, S9, S10 | $Ir_{21}Pd_{21}Pt_{19}Rh_{17}Ru_{22}$ |
| Figure S4a | $Ir_{14}Pd_{13}Pt_{21}Rh_{9}Ru_{43}$* |

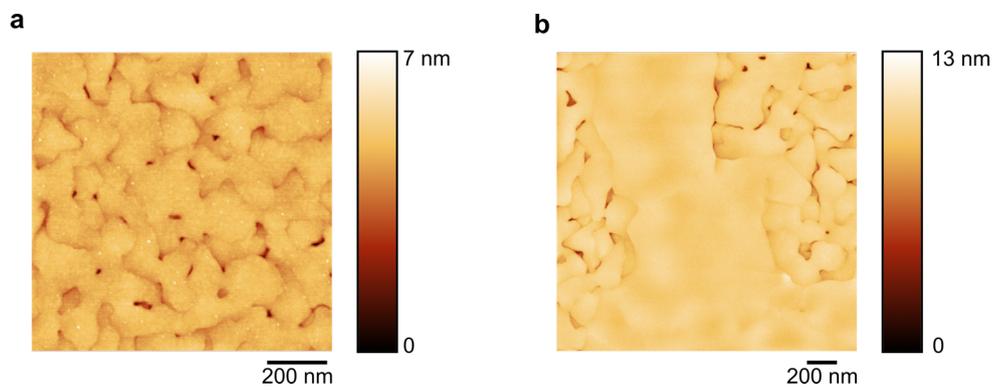

**Figure S1.** AFM surface topography images of (a) 60 nm thick epitaxial CCSS film grown on 5 nm Pt buffer and (b) 100 nm thick epitaxial CCSS film grown on 40 nm Pt buffer.

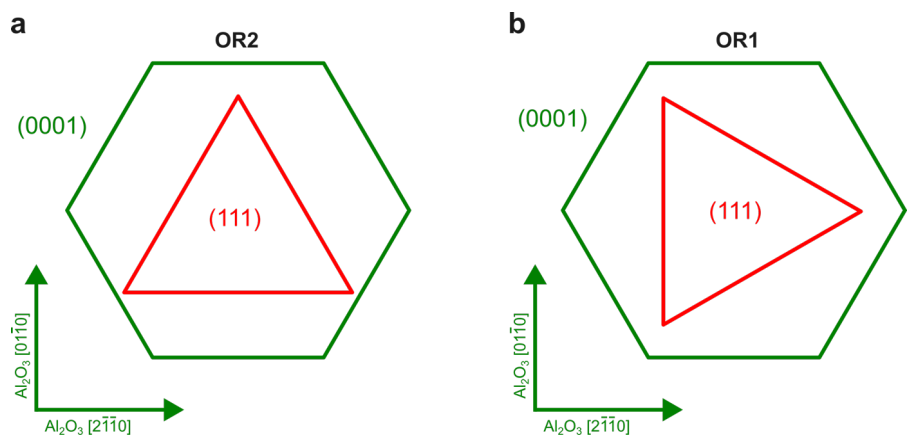

**Figure S2.** Illustrative sketch depicting the epitaxial growth relations (a) OR2 and (b) OR1 observed between (111) plane of the CCSS film in red and (0001) plane of the sapphire substrate in green.

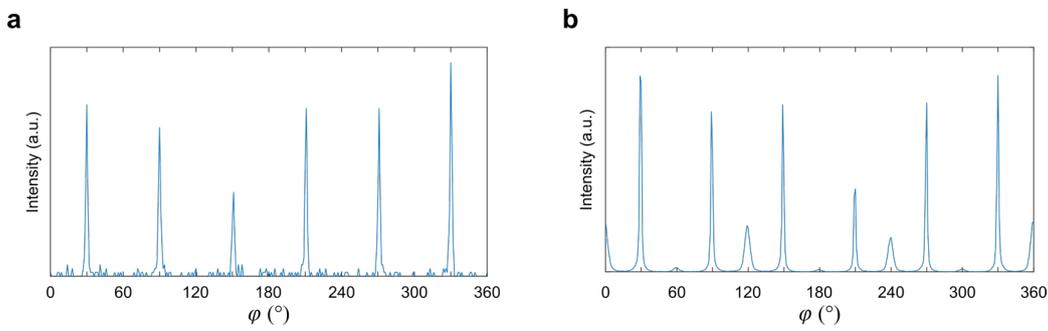

**Figure S3.** XRD results of {220} φ-scans of the Pt buffer layers of (a) the 60 nm and (b) the 100 nm epitaxial CCSS film sample. The Pt buffer layer grows with OR2 epitaxial relation on $Al_2O_3(0001)$ in case of the 60 nm epitaxial CCSS film and a combination of OR1 and OR2 epitaxial relations in case of the 100 nm epitaxial CCSS film.

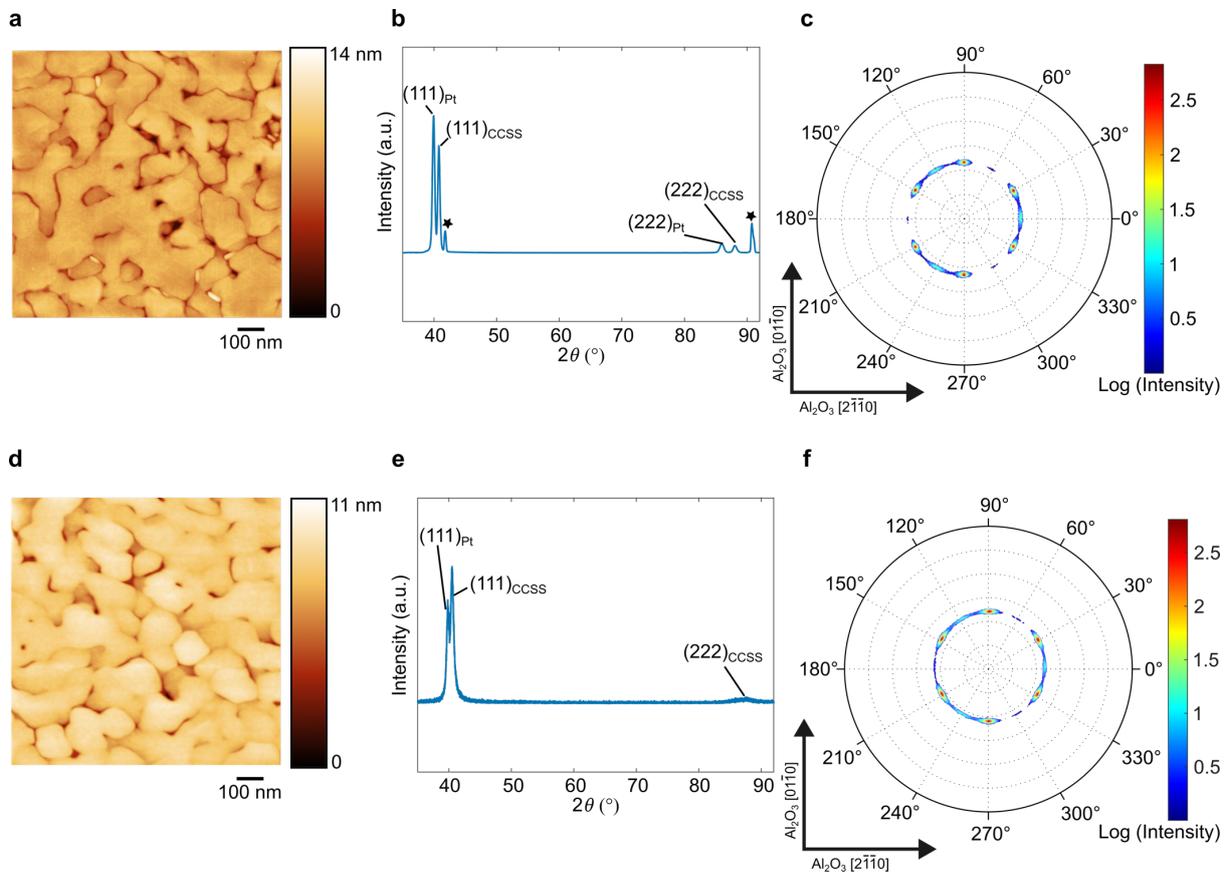

**Figure S4.** Demonstration of epitaxial growth over a wide CCSS composition range using a Ru-rich and a Ru-free CCSS system. (a-c) AFM surface topography, X-ray diffractogram and {220} pole figure of a 100 nm thick $Ir_{14}Pd_{13}Pt_{21}Rh_9Ru_{43}$ CCSS film deposited on 40 nm thick Pt buffer layer at 873 K. (d-f) AFM surface topography, X-ray diffractogram and {220} pole figure of a 100 nm thick near equiatomic Ir-Pd-Pt-Rh film deposited on 40 nm thick Pt buffer layer at 873 K. AFM surface topography of both CCSS films depicts flat discontinuous grains. X-ray diffractogram of both films show {111} and {222} peaks of Pt buffer and CCSS film along with sapphire substrate peaks in (b), indicated by ★. The {220} pole figures of the CCSS films exhibit both OR1 and OR2 epitaxial relations with $Al_2O_3(0001)$ substrate.

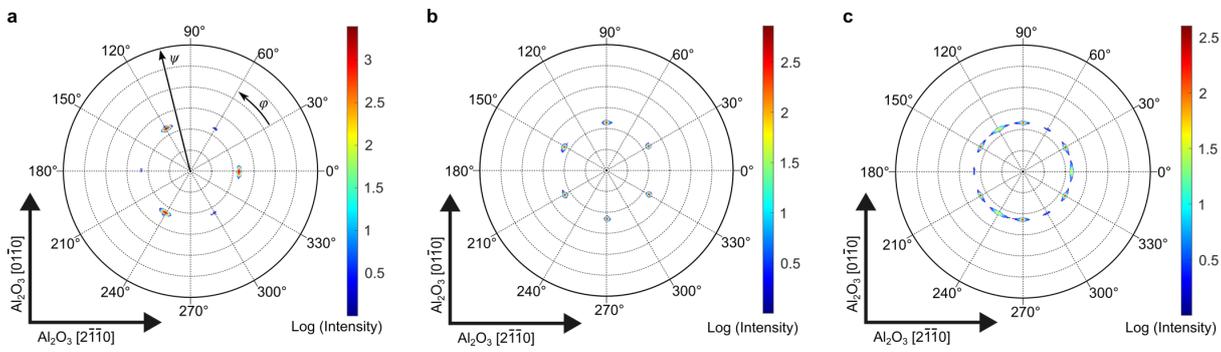

**Figure S5.** Complete {220} pole figures of (a) 60 nm Pt film (b) 60 nm CCSS film grown on 5 nm Pt buffer layer and (c) 100 nm CCSS film grown on 40 nm Pt buffer layer.

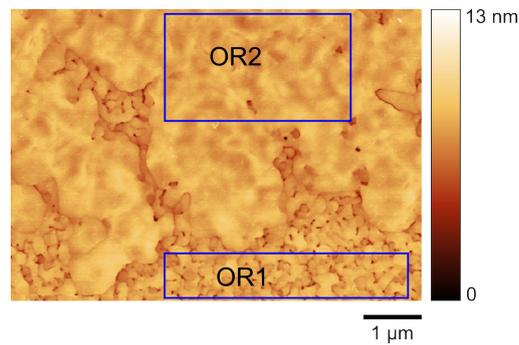

**Figure S6.** AFM surface topography map of the 100 nm epitaxial CCSS film region used for correlative investigations. The blue boxes depict the regions analyzed for determining RMS roughness of OR1 and OR2 grains. The OR1 region has a roughness of 0.75 nm and the OR2 region has a roughness of 0.53 nm.

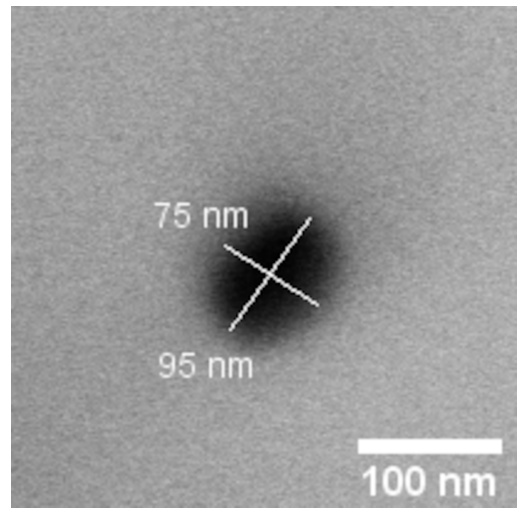

**Figure S7.** SEM image of the tip opening of the single barrel quartz nanopipette used for SECCM measurements. A diameter ($d$) of 85 nm is obtained using $d = \sqrt{ab}$, where $a$ and $b$ are the semi-major and semi-minor axes of the nanopipette.

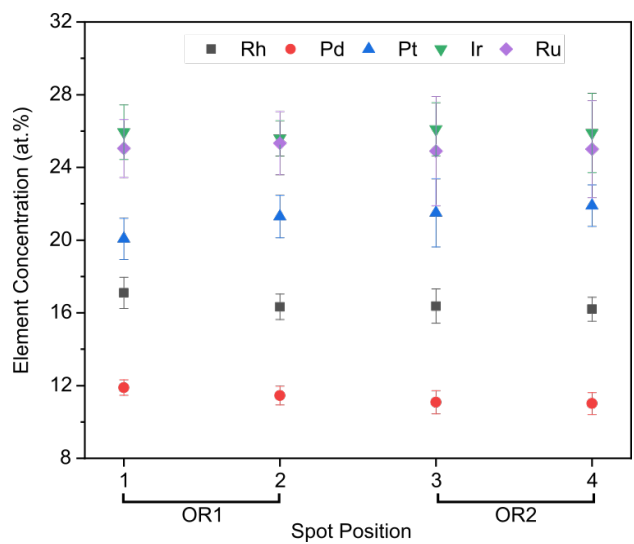

**Figure S8.** Plot summarizing the surface composition of 100 nm thick epitaxial CCSS film grown on 40 nm thick Pt buffer determined by Auger electron spectroscopy. Four exemplary spots, two in OR1 and two in OR2 regions, were used for quantification. The film composition is uniform across both regions of different epitaxial growth relations. The error bars denote fitting error. The observed deviation from the STEM-EDS composition ($Ir_{21}Pd_{21}Pt_{19}Rh_{17}Ru_{22}$) is due to carbon build-up on the surface during the measurement. Carbon deposition occurs due to the cracking of hydrocarbon molecules, which likely originate from the conductive silver paste used for prior EBSD characterization. Since Pd (333 eV) and Rh (305 eV) were analyzed at lower kinetic energies than Ir (1909 eV), Pt (2048 eV), and Ru (2256 eV), Pd and Rh signals are more strongly attenuated by surface carbon, leading to an underestimation of Pd and Rh.

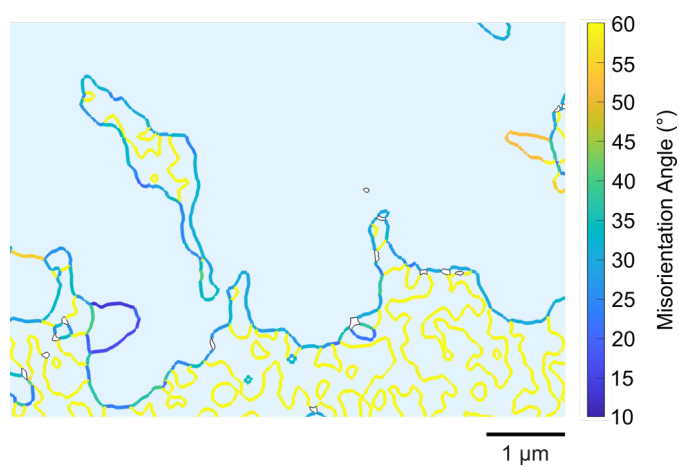

**Figure S9.** Map depicting the misorientation angle of the grain boundaries present in the region used for the correlative AFM-SECCM-EBSD investigation in Figure 5 of the manuscript. The grains are shown as the light blue background. The grain boundaries are densely present in OR1 region. The boundaries, highlighted in yellow, have a misorientation of [111] 60°, which makes them {112}Σ3 incoherent twin boundaries.

# 1. Co-localization approach for correlative characterization by AFM-SECCM-EBSD

For correlative characterization across SECCM, AFM, and EBSD, fiducial markers are required on the film surface to enable reliable sample navigation, alignment and relocation of regions of interest. In this work, we used micro-indents as surface markers. Unlike lithographically defined markers, micro-indents do not introduce chemical contamination and are readily detectable in both optical microscopy (integrated with SECCM and AFM) and electron microscopy (EBSD). The indents can also be created in a sequential series using the micro-indenter, establishing well-defined relative spatial coordinates on the film surface.

An L-shaped array of indents (Figure S10a) was created at the sample edge to define the global X–Y reference frame during mounting. After aligning to this reference in the SECCM set-up, we navigated to a measurement region, using a pre-defined indent with known coordinates relative to the L-shaped marker.

The SECCM tip was landed beside the indent, and a coarse reference scan was acquired, covering an area of 30x30 µm$^2$ with a hopping distance of 1 µm. This coarse scan provides the topographic map of the region, which helps to determine the probe location relative to the indent (Figure S10b) and check for trends in the current map (Figure S10c) to identify any influence of the indentation on the surrounding film surface. To optimize experiment time while probing extended areas, the electrochemical protocol just included the approach at -0.45 V vs. Ag/AgCl (3.4 M KCl) and one cyclic voltammetry scan between -0.160 V and -0.810 V vs. Ag/AgCl (3.4 M KCl) at 2 V s$^{-1}$.

The topography and current map of the coarse reference scan (Figure S10b, c) allowed to estimate the center of the indent relative to the center of the scan area. Following the coarse reference scan, the intended high-resolution fine scan was performed after translating the piezo stage (ΔX: 40 µm, ΔY: -3 µm) from the coarse scan center. This calibrated displacement along with the coarse reference scan dataset helped to define the precise location of the high-resolution SECCM scan relative to the indent center. This spatial reference was then used to subsequently relocate the scanned region for AFM and EBSD characterization.

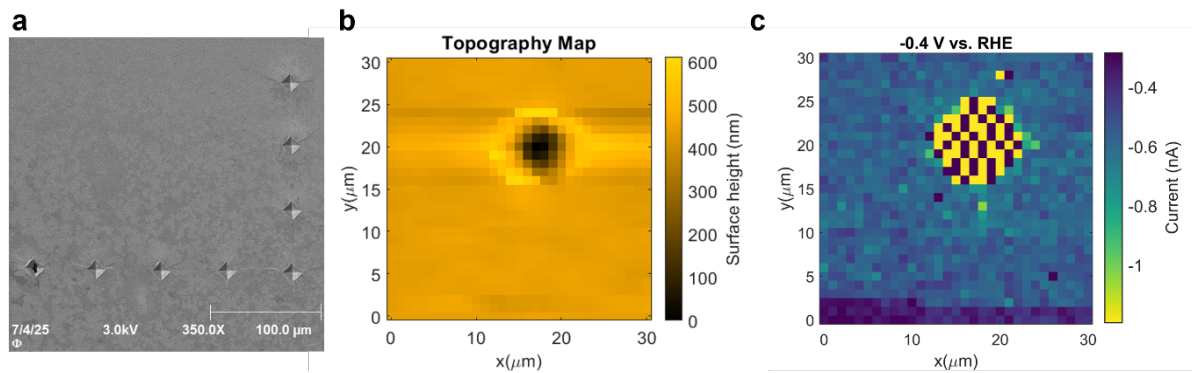

**Figure S10.** (a) SEM image of an L-shaped group of indents used for sample alignment reference. (b) SECCM topography map from the coarse reference scan acquired in the region of the targeted indent. (c) Corresponding current map recorded at -0.4 V vs. RHE around the targeted indent.